\documentclass[prd,preprint]{revtex4}

\newcommand\Lr{\rm L}

\usepackage{xcolor}

\usepackage[utf8]{inputenc}

\usepackage{latexsym}
\usepackage{epsfig}
\usepackage{graphicx,subfigure}
\usepackage[normalem]{ulem}
\usepackage{color}
\usepackage{xcolor}
\usepackage{multirow}
\usepackage{hyperref}
\usepackage{numprint}
\usepackage{eucal}
\usepackage{amsmath}
\usepackage{amssymb}

	\usepackage{amsmath}
	\usepackage{amsfonts}
  	\usepackage{amssymb}
	\usepackage{makeidx}
	\usepackage{amsfonts}
	\usepackage[ansinew]{}
	\usepackage[usenames,dvipsnames]{pstricks}
	\usepackage{epsfig}

    \usepackage{graphicx}

	\usepackage{float}

\newcommand{\be}{\begin{equation}}
\newcommand{\ee}{\end{equation}}
\newcommand{\bea}{\begin{eqnarray}}
\newcommand{\eea}{\end{eqnarray}}
\newcommand{\ba}{\begin{align}}
\newcommand{\ea}{\end{align}}

\makeindex

\begin{document}

\title{Testing the consistency of cosmological models with multimessenger astronomy}

\author{Antonio Enea Romano}

\affiliation{Instituto de Fisica,Universidad de Antioquia,A.A.1226, Medellin, Colombia}
\affiliation{ICRANet, Piazza della Repubblica 10, I--65122 Pescara, Italy}


\begin{abstract}
     We show that within the framework of the Friedmann cosmological model it is possible to derive general multimessenger consistency conditions, between the gravitational wave (GW) and the electromagnetic wave (EMW) luminosity distances.
    We first study the effects of the flatness assumption often used in the estimation of cosmological parameters from GWs observations, deriving a general relation for the curvature parameter,  uniquely in terms of the flatness biased GW and the EMW luminosity distances, and  a  multimessenger consistency relation for the Friedmann model, valid  independently of the dark energy equation of state.  
    We then derive a general
     multimessenger consistency relation for the cosmological constant, which is independent of any flatness assumption  made in the analysis of observational data. 
\end{abstract}


\maketitle

\section{Introduction}
With the detection of gravitational waves (GWs) \cite{LIGOScientific:2016aoc} by the Laser Interferometer Gravitational Wave Observatory (LIGO) and Virgo, the era of gravitational multi-messenger astronomy has started. 
These new set of observations combined with other cosmological probes allow to perform new tests of the standard cosmological model based on the Friedmann equations. The recent analysis of large scale structure data \cite{DESI:2024mwx} has shown preference for dynamical dark energy models, but  this conclusion can partially depend on the choice of parametrization \cite{Gialamas:2024lyw,Wolf:2025jlc} adopted for dark energy. 
In order to avoid biases due to the choice of different parametrizations for the equation of state of dark energy, it is useful to introduce model independent tests, which can be applied directly to observational quantities.

We show that the combination of GW and EMW luminosity distances observations allow to probe the curvature of the Universe independently of the equation of state of dark energy and of other cosmological density parameters.
We then derive a general multimessenger consistency condition for the Friedmann model, in terms of the GW and EMW luminosity distances, which is independent of the equation of state of dark energy and of the value of the density parameters.

In order to test the dynamical nature of dark energy, we also derive a multimessenger consistency test for the cosmological constant, involving only  the GW and EMW luminosity distances, and no other cosmological parameter.

\section{Electromagnetic waves luminosity distance}
In a Friedmann Universe \cite{1922ZPhy...10..377F} the electromagnetic luminosity distance is given by
\begin{equation}\label{d_L}
d^{EM}_{L}(z)= \frac{a_o c(1+z)}{H_0 \sqrt{-\Omega_k}}\sin{\left( 
\sqrt{-\Omega_k}\int_0^z{\mathrm{d}z'\frac{H_0}{H(z')}}\right)},
\end{equation}
where the Hubble parameter $H(z)$ is given by the Friedmannn equation, 
\be
H(z)^2= H_0^2\biggl\{\Omega_{m} (1+z)^3+\Omega_{
k}(1+z)^2
+\Omega_{DE}\exp{\left[3\int_0^z
\frac{1+w(z')}{1+z'}\mathrm{d}z'\right]}\biggr\}, \label{Hz}
\ee
in which $w(z)$ denotes the equation of state of dark energy and the density parameters satisfy the relation $\Omega_{DE}=1-\Omega_m-\Omega_k$.
For a flat Universe we have 
\be
d^{EM}_{L}(z,\Omega_k=0)=d^{EM}_{L,f}(z)= a_o c(1+z) \int_0^z \frac{\mathrm{d}z'}{H(z')}.
\ee

\section{Gravitational waves luminosity distance}
Assuming general relativity \cite{1916AnP...354..769E}, the leading order calculation of the amplitude of GWs emitted by a binary coalescence yields \cite{1918SPAW.......154E,1951ctf..book.....L,1963PhRev.131..435P,1986Natur.323..310S}
\be
\gamma_{ij,e}\propto \frac{1}{r } {M_c}^{5/3} f_e^{2/3} \,, \label{hGR}
\ee
where $M_c$ is the chirp mass, and $f_e$ is the emitted GW frequency.
Assuming a massless graviton, the  cosmological expansion implies 
\be
\frac{\gamma_{ij,o}}{\gamma_{ij,e}}=\frac{a_e}{a_o} \label{ha}\,,
\ee
and the redshift is defined as
\be
(1+z)=\frac{a_o}{a_e}\label{za}\,.
\ee
The emitted and observed frequencies, $f_e$ and $f_o$, are related by 
\be
f_e=(1+z)f_o \,,
\ee
which combined with eq.(\ref{hGR}) and eq.(\ref{ha}) give \cite{Finn:1992xs}
\be
\gamma_{ij,o}\propto \frac{1}{r}\frac{a_e}{a_o} {M_c}^{5/3} (1+z)^{2/3}f_o^{2/3}= \frac{1}{d^{\rm GW}_{\Lr}(z)} {M_c}^{5/3} (1+z)^{5/3}f_o^{2/3}=\frac{1}{d^{\rm GW}_{\Lr}(z)} {\mathcal{M}_c}^{5/3} f_o^{2/3} \,. \label{hGRo}
\ee
In the above equation we have introduced the definition of the redshifted chirp mass $\mathcal{M}_c=(1+z)M_c$  and of the gravitational luminosity distance as
\be
d^{\rm GW}_{\Lr}(z)=a_o r(z) (1+z) \,.\label{dGW}
\ee
\section{The effects of flatness assumption in the analysis of GWs data}
In following we will assume that the comoving distance $r(z)$ appearing in eq.(\ref{dGW}) is given by
\be
 r(z)=c \int_0^z \frac{\mathrm{d}z'}{H(z')} \,.\label{rflat}
\ee
Note that the above formula is only valid for a flat Universe, while in general $r(z)$ would have an additional functional dependence on the curvature. Flatness is assumed in the inference of cosmological parameters from GWs data \cite{LIGOScientific:2025jau}, including the GW-EMW distance ratio, and for consistency  we have assumed it in eq.(\ref{rflat}), while from a theoretical point of view $r(z)$ should also include the extra curvature dependence, giving the well known general relativity prediction  $d^{\rm GW}_{\Lr}(z)=d^{\rm EM}_{\Lr}(z)$. 
The quantity $d^{\rm GW}_{\Lr}(z)$ appears in the denominator of the observed GW amplitude as a result of the mathematical manipulation leading to the definition of redshifted chirp mass. In the following we will assume $a_o=1$ and define $D(z)=(H_0/c)(1+z)^{-1}d_L(z)$, which we will use for both the GW and EMW luminosity distance.

For the GW luminosity distance we have
\be
\frac{d}{dz}D_{GW}=\frac{H_0}{H(z)} \,,\label{DGWpflat}
\ee
while for the EMW luminosity distance we can obtain this relation
\be
\left[\frac{d}{dz}D_{EM}\right]^2=\frac{H_0^2 \left[\Omega_k D_{EM}(z)^2+1\right]}{H(z)^2} \,, \label{DEMp}\,
\ee
which has been derived using trigonometric identities to re-write the derivative in terms of $D_{EM}(z)$.
The above equations are valid of any form of dark energy. 
Eq.(\ref{DGWpflat}) is very important, since it allows to obtain $H(z)$ from GWs observations, which is the key to obtain multimessenger relations.

\subsection{Multimessenger estimation of curvature}
 Combining eq.(\ref{DEMp}) and eq.(\ref{DGWpflat}), we obtain a multimessenger estimation of the curvature parameter
\be
\Omega_{k}=\frac{1}{D_{EM}(z)}\left\{ {\left[\frac{D_{EM}'(z)}{D_{GW}'(z)}\right]}^2-1 \right\} \,. \label{Okmm}
\ee
The above equation allows to estimate the curvature from the difference between the two luminosity distances. As expected, when $D_{EM}(z)=D_{GW}(z)$, we have $\Omega_k=0$. 

The difference between the GW and EMW luminosity distance could potentially also be due to the effects of gravity modification \cite{Romano:2023xal,Romano:2023ozy}, so  eq.(\ref{Okmm}) provides also a test to distinguish between the effects of cosmic curvature and gravity modification on the GW luminosity distance.

\subsection{Friedmann model multimessenger consistency condition}
Since $\Omega_k$ is constant in a homogeneous Universe, we can obtain another more general consistency condition by taking the derivative of eq.(\ref{Okmm}), giving
\be
\mathcal{C}_{F}(z)=
D_{EM}(z) D_{EM}''(z) D_{GW}'(z)-D_{EM}(z) D_{EM}'(z) D_{GW}''(z)-D_{EM}'(z)^2 D_{GW}'(z)+D_{GW}'(z)^3=0 \,.
\ee
This equation holds independently of the equation of state of dark energy, which is consistent with the fact that eq.(\ref{Okmm}) holds independently of the equation of state of dark energy.

\section{General Friedmann multimessenger consistency condition}
In the previous sections we have defined the GW luminosity distance using eq.(\ref{rflat}), following the flatness assumption made in the GWs data analysis.
From a general theoretical point of view we have $d^{EM}_{L}(z)=d^{GW}_{L}(z)$, which we can use to obtain a more general consistency condition.
For a Universe with dark energy in the form of a cosmological constant, we have 
\be
\Omega^{GW}_{m}=\frac{1-\left[\Omega_k z (z+2)+1\right] D_{GW}'(z)^2+\Omega_k D_{GW}(z)^2}{z \left(z^2+3 z+3\right) D_{GW}'(z)^2} \,.\label{Omgwk}
\ee
We can obtain an expression for $\Omega_k$ by solving eq.(\ref{DEMp}), giving \cite{Clarkson:2007bc,Heinesen:2026wwy} 
\be
\Omega_k(z)=\frac{H(z)^2 D_{EM}'(z)^2-H_0^2}{\left[H_0^2 D_{EM}(z)\right]^2} \,. \label{Okdem}
\ee
and after replacing  $\Omega_k$ using eq.(\ref{Okdem}) 
we get a general multimessenger formula of $\Omega_m$
\be
\Omega^{MM}_{m}=-\frac{\left(D_{GW}^2-z (z+2) D_{GW}'^2\right) \left(H^2 D_{EM}'^2-H_0^2\right)+H_0^2 D_{EM}^2 \left(1-D_{GW}'^2\right)}{H_0^2 z (z (z+3)+3) D_{EM}^2 D_{GW}'^2}
\ee
This quantity should be constant, so taking the derivative we get a general multimessenger consistency condition for the cosmological constant
\bea
\mathcal{C}_{\Lambda}&=&2 D_{GW}' [z (z+2) D_{GW}''-2 H_0^2 D_{EM} D_{EM}' (D_{GW}'^2-1)+ \\
&&+(z+1) D_{GW}' 
-D_{GW}] (H_0^2-H^2 D_{EM}'^2)+ \\
&&+2 H D_{EM}' [D_{GW}^2-
 z (z+2) D_{GW}'^2] 
(H D_{EM}''+D_{EM}' H')-2 H_0^2 D_{EM}^2 D_{GW}' D_{GW}''=0 \,.\label{CL}
\eea
This consistency condition combines GW and EMW observables and is independent of density parameters


\section{Conclusions}
We have derived different multimessenger consistency conditions for the Friedmann cosmological model, involving the GW and EMW luminosity distances. First we have studied the effects of the flatness assumption adopted in the analysis of GWs data, and derived a consistency relation for the Friedman model, independent from the form  of dark energy and other cosmological density parameters. Under this same assumption we have also derived an expression for the curvature, allowing to distinguish between the effects of gravity modification and curvature on the GW-EMW distance ratio constraints derived assuming flatness for GWs observations analysis.

We have then derived a more general multimessenger consistency for the cosmological constant,  expressed directly in terms of observational quantities, independent of any assumption of flatness adopted in analyzing observational data. 
This general multimessenger consistency condition for the cosmological constant can be applied directly to observational data, independently of the assumptions for the curvature made in the GWs data analysis. In the future it will be interesting to apply this consistency conditions to the results of observational data analysis, in order to set multimessenger constraints on curvature using GW-EMW distance ratio constraints derived assuming flatness. 
Another interesting application could be a consistency test of  the cosmological constant using multimessenger observations.

\section{Acknowledgements}
This work was supported by the UDEA project No. 2023-63330.

\bibliographystyle{h-physrev4}
\bibliography{mybib}
\end{document}